\documentclass[Conference]{IEEEtran}

\IEEEoverridecommandlockouts
\usepackage{color}
\usepackage{graphicx}
\usepackage{amsmath}
\usepackage{amssymb}
\usepackage{algorithm}
\usepackage{algorithmic}
\usepackage{amsmath}
\usepackage{multirow}
\usepackage{booktabs}
\usepackage{array}
\usepackage{amsthm}
\usepackage{stfloats}
\usepackage{caption}
\usepackage{subfigure}
\usepackage{bm}
\usepackage{booktabs}
\usepackage{setspace}

\newcommand{\be}{\begin{equation}}
\newcommand{\ee}{\end{equation}}
\newcommand{\bea}{\begin{eqnarray}}
\newcommand{\eea}{\end{eqnarray}}
\newcommand{\ba}{\begin{array}}
\newcommand{\ea}{\end{array}}

\title{Symbol-Level Precoding Design for Intelligent Reflecting Surface Assisted Multi-user MIMO Systems
\thanks{$^{\ast}$ Corresponding author.}
\thanks{This paper is supported by the National Natural Science Foundation of China (Grant No. 61971088, 61671101, 61601080, and 61761136019) and the Natural Science Foundation of Liaoning Province (Grant No. 20180520019).}}
\author{\IEEEauthorblockN{Rang Liu$^{\dag}$, Hongyu Li$^{\dag}$, Ming Li$^{\dag}$$^{\ast}$, and Qian Liu$^{\ddag}$
\vspace{-0.0 cm} }\\
\IEEEauthorblockA{$^{\dag}$School of Information and Communication Engineering   \\  Dalian University of Technology, Dalian, Liaoning 116024, China \\ E-mail: \texttt{\{liurang, hongyuli\}@mail.dlut.edu.cn, mli@dlut.edu.cn} } 

\IEEEauthorblockA{$^{\ddag}$School of Computer Science and Technology \\  Dalian University of Technology, Dalian, Liaoning 116024, China \\ E-mail: \texttt{qianliu@dlut.edu.cn} }}

\begin{document}

\maketitle
\pagestyle{empty}
\thispagestyle{empty}

\begin{abstract}
Intelligent reflecting surface (IRS) has emerged as a promising solution to enhance wireless information transmissions by adaptively controlling prorogation environment. Recently, the brand-new concept of utilizing IRS to implement a passive transmitter attracts researchers' attention since it potentially realizes low-complexity and hardware-efficient transmitters of multiple-input single/multiple-output (MISO/MIMO) systems.
In this paper we investigate the problem of precoder design for a low-resolution IRS-based transmitter to implement multi-user MISO/MIMO wireless communications.
Particularly, the IRS modulates information symbols by varying the phases of its reflecting elements and transmits them to $K$ single-antenna or multi-antenna users.
We first aim to design the symbol-level precoder for IRS to realize the modulation and minimize the maximum symbol-error-rate (SER) of single-antenna receivers.
In order to tackle this NP-hard problem, we first relax the low-resolution phase-shift constraint and solve this problem by Riemannian conjugate gradient (RCG) algorithm.
Then, the low-resolution symbol-level precoding vector is obtained by direct quantization.
Considering the large quantization error for 1-bit resolution case, the branch-and-bound method is utilized to solve the 1-bit resolution symbol-level precoding vector.
For multi-antenna receivers, we propose to iteratively design the symbol-level precoder and combiner by decomposing the original large-scale optimization problem into several sub-problems.
Simulation results validate the effectiveness of our proposed algorithms.
\end{abstract}


\maketitle

\section{Introduction}
The fifth generation (5G) wireless communication is arriving, which is characteristic of high data rate, dense connections and low latency.
Various technical approaches have emerged to improve the wireless network performance, which encompasses massive multiple-input multiple-output (MIMO), millimeter wave (mmWave) communications, and ultra-dense network.
All these technologies have to adapt to the complicated and time-varying wireless communication environment and ensure the wireless information transmissions in face of channel fading.
On the contrary, the emerging intelligent reflecting surface (IRS) technology turns to change the channel environment by utilizing some very hardware-efficient reflecting elements.
Therefore, IRS has attracted significant attention in the past several months since it can enhance the quality of communications in an efficient and green fashion \cite{Renzo 19}.

IRS is a planar array composed of some passive, low-cost reflecting elements, which adjust the phase of the incident electromagnetic wave and reflect it without any power consumption.
These reflecting elements are made up of some hardware-efficient devices, e.g.  positive-intrinsic-negative (PIN) diodes and phase shifters.
With the aid of recent development of micro-electrical-mechanical systems (MEMS) and meta-materials, the reflecting surfaces can be reconfigured in real-time, which makes IRS more appealing than the conventional relay and backscatter communications \cite{Basar survey}.
Therefore, IRS provides supplementary links to combat the blockage caused by buildings, trees or cars, and improves the performance by changing the channel environment.
Besides, these programmable and controllable reflecting elements are lightweight, which enables them being attached to the buildings or some mobile objects.

Considering the advantages mentioned above, many researches have been springing up in recent months to exploit the IRS in existing wireless communication networks.
The simple but representative single-user systems have been extensively studied, e.g. the ergodic spectral efficiency maximization problem \cite{Han 08746155}, the power minimization problem with low-resolution IRS \cite{Wu 08683145}, the signal power maximization problem \cite{Wu 08647620}, and the secrecy rate maximization problem \cite{Cui IWCT}, \cite{Shen ICT}.
Then, the utilization of IRS in the multi-user systems was further investigated to enhance system performance by improving the propagation environment \cite{Guo 190507920}-\cite{Chen 08742603}.
By properly tuning each reflecting element of IRS, the multi-user interference (MUI) is suppressed and the performance loss due to the signal attenuation and scattering is compensated.

In the applications mentioned above, the IRS only reflects the incident signals, which are already modulated and precoded by a transmitter.
In particular, the transmitter of a MIMO system requires a number of power-hungry radio frequency (RF) chains and digital-to-analog converters (DACs) to process the transmitted signals.
In order to realize a low-complexity and energy-efficient system, the concept of utilizing IRS as a transmitter to serve one single-antenna receiver was proposed in \cite{Basar IRS-AP 2019}.
The IRS changes each reflecting element to modulate and transmit information symbols by exploiting an unmodulated carrier signal, which is generated from a nearby RF signal generator.
In this way, the virtual MIMO system is implemented with only one RF chain.

Since the information is modulated and transmitted by varying each reflecting element of IRS, the IRS design problem is similar to the symbol-level precoding problem \cite{Masouros ITWC 09}-\cite{Liu ISPL 17}, where the transmitted precoding vector changes with the symbol vector.
Therefore, based on the findings in symbol-level precoding, the IRS can serve multiple users and exploit MUI as a green power to enhance the information transmissions.
Moreover, each reflecting element of IRS usually can only change its phase.
Thus, the IRS design problem is actually a constant envelope precoding problem as in \cite{Liu ISPL 17}, where the infinite resolution phases are exploited.
Considering the high energy consumption and hardware complexity due to the infinite resolution phases, the low-resolution reflecting elements are usually employed in practical IRSs.
Furthermore, existing works \cite{Masouros ITWC 09}-\cite{Liu ISPL 17} of symbol-level precoding only consider the single-antenna receiver case.
To the best of our knowledge, the symbol-level precoding for multi-antenna receivers has not been studied, which motivates us to investigate this problem.

In this paper, we aim to design the low-resolution symbol-level precoder for a IRS-based transmitter to minimize the maximum symbol-error-rate (SER) of all receivers.
Particularly, we consider that the IRS with $N$ low-resolution reflecting elements acts as a transmitter to serve $K$ single-antenna or multi-antenna users.
The optimization problem is to minimize the maximum SER with the low-resolution phase constraint for the precoder.
To tackle this NP-hard problem, we first attempt to obtain the precoder with infinite resolution and then quantize it into discrete phase values.
Since the constant modulus constraint forms a manifold, we attempt to smooth the objective function and solve it in the Riemannian space by the popular Riemannian conjugate gradient (RCG) algorithm.
For the 1-bit resolution IRS, the quantization error is too large and the performance degrades significantly.
Thus, we further use the branch-and-bound method to solve the mixed integer nonlinear programming (MINLP) problem for the 1-bit resolution precoder design, which provides satisfactory performance.
For multi-antenna receivers, the low-resolution precoder and the combiner are designed in an iterative fashion.
When the combiner is fixed, we decompose the large-scale optimization problem into some sub-problems, where each precoder is designed in the same way as the single-antenna case.
When the precoder is fixed, the combiner is obtained by converting the optimization problem into convex sub-problems.
Simulation results demonstrate the effectiveness of our proposed algorithms.

The following notations are used throughout this paper.
Boldface lower-case and upper-case letters indicate column vectors and matrices, respectively.
$(\cdot)^T$, $(\cdot)^*$, and $(\cdot)^H$ denote the transpose, the conjugate, and the transpose-conjugate operations, respectively.
$\mathbb{C}$ denotes the set of complex numbers.
$| a |$ and $\| \mathbf{a} \|$ are the magnitude of a scalar $a$ and the norm of a vector $\mathbf{a}$, respectively.
$[a]$ and $\angle{a}$ are the round of a scalar $a$ and the angle of complex-valued $a$, respectively.
$\mathfrak{R}\{\cdot\}$ and $\mathfrak{I}\{\cdot\}$ denote the real part and the imaginary part of a complex number, respectively.
Finally, we adopt a Matlab-like matrix indexing notation: $\mathbf{A}(i,j)$ denotes the element of the $i$-th row and the $j$-th column of matrix $\mathbf{A}$, $\mathbf{a}(i)$ denotes the $i$-th element of vector $\mathbf{a}$.


\section{System Model and Problem Formulation}
\vspace{0.2 cm}
\begin{figure}[!t]
\centering
\includegraphics[height=2.3 in]{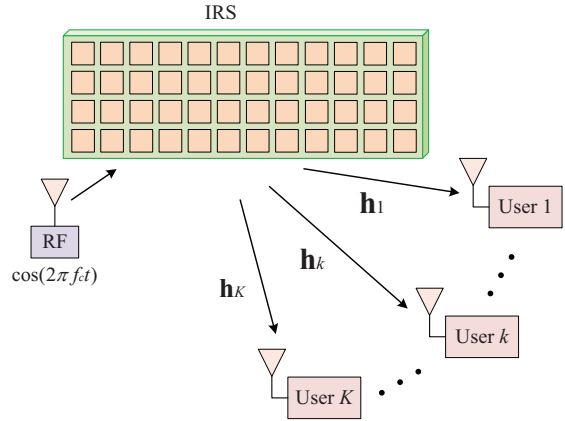}
\vspace{-0.1 cm}
\caption{The IRS assisted MU-MISO system.}
\label{fig:system model}
\vspace{-0.3 cm}
\end{figure}

We first consider a multiuser multiple-input single-output (MU-MISO) system as shown in Fig. 1, which utilizes IRS as a transmitter.
The IRS has $N$ passive reflecting elements and attempts to serve $K$ single-antenna users by modulating information symbols to the high frequency carrier signal $\cos(2\pi f_\mathrm{c}t)$ generated from a nearby RF signal generator.
We assume the RF signal generator is very close to the IRS and the channel fading does not affect the transmission.
In this case, information symbols are modulated by adjusting the phases of IRS's reflecting elements.
Therefore, the IRS design is in a symbol-by-symbol fashion and similar to the symbol-level precoding technique, which is a hot topic recently and can enhance the information transmissions by exploiting MUI with known desired symbol vector and channel state information (CSI).

Assume that $K$ independent symbols $s_1,\ldots,s_K$, are selected form the $M$-phase-shift-keying (PSK) modulation symbols.
Then, the symbol vector $\mathbf{s}=[s_1,\ldots,s_K]^T$ has $M^K$ different combinations.
For different desired symbol vectors $\mathbf{s}_m, m = 1,\ldots,M^K$, the IRS changes its reflecting elements accordingly to implement the MISO wireless communications.
Correspondingly, we denote the values of all the reflecting elements of IRS for transmitting symbol vector $\mathbf{s}_m$ as ${\bm \theta}_m=[\theta_{m,1},\ldots,\theta_{m,N}]^T$.
Since the energy consumption and hardware complexity are proportional to the resolution of each reflecting element, the infinite resolution reflecting elements are impractical in the real world.
Therefore, we consider the low-resolution reflecting elements, which only take finite discrete phase values.
Assuming the phases are controlled by $B$ bits, the set of these discrete phase values is
\begin{equation}
\mathcal{F} \triangleq \left\{e^{j\frac{2\pi b}{2^B}} | b= 0,1,\ldots,2^B-1\right\},
\end{equation}
and
\begin{equation}
\bm{\theta}_m(n) \in \mathcal{F}, n=1,\ldots, N.
\end{equation}

Therefore, when the symbol vector to be transmitted is $\mathbf{s}_m$, the symbol-level precoder of IRS is $\bm{\theta}_m$ and the received signal at the $k$-th user can be written as
\begin{equation}
r_{k,m} = \sqrt{\frac{P}{N}}\mathbf{h}_k^H \bm{\theta}_m + n_k, \forall k,
\end{equation}
where $P$ is the total transmit power, $\mathbf{h}_k \in \mathbb{C}^{N \times 1}$ is the channel vector from IRS to the $k$-th user, and $n_k \sim \mathcal{CN}(0,\sigma_k^2)$ is the additive white Guassian noise (AWGN) at the $k$-th user.
In this paper, we assume all users' CSI is perfectly known at IRS, which can be obtained by some channel estimation approaches.
\begin{figure}[!t]
\centering
\subfigure[An example.]{
\begin{minipage}{4.2 cm}
\centering
\includegraphics[height=3.65 cm]{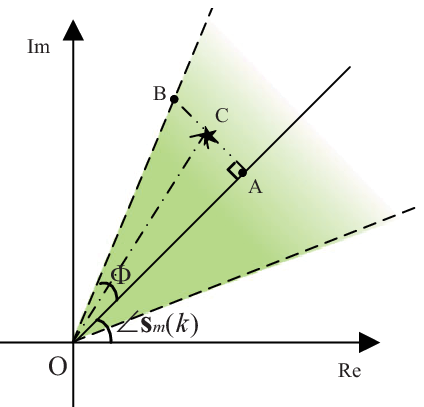}
\vspace{0.3 cm}
\label{fig:CR1}
\end{minipage}}
\subfigure[After rotating Fig. 2(a).]{
\begin{minipage}{4.2 cm}
\centering
\includegraphics[height=3.65 cm]{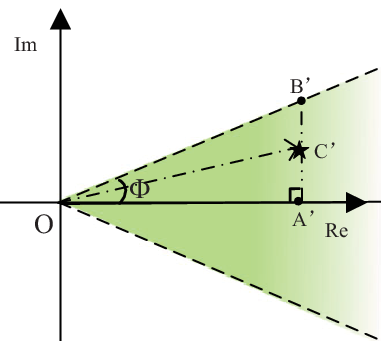}
\vspace{0.3 cm}
\label{fig:CR2}
\end{minipage}}
\caption{The symbol-level precoding design for 8-PSK signals.}
\label{fig:CR}
\vspace{-0.4 cm}
\end{figure}

From \cite{Masouros ITWC 09}-\cite{Liu ISPL 17}, we learn that the MUI in the MU-MISO systems can be utilized as constructive power to enhance the information transmissions when the IRS (i.e. the transmitter) exploits $\mathbf{s}_m$ and the knowledge of CSI to design the symbol-level precoder $\bm{\theta}_m$.
In order to illustrate how to utilize MUI, we take the 8-PSK modulated signals as an example and plot the decision region of symbol $(1/\sqrt{2},j/\sqrt{2})$ in Fig. \ref{fig:CR1}, where point $C$ denotes the received noise-free signal $\mathbf{h}_k^H\bm{\theta}_m$, point $A$ denotes the projection of point $C$ in the direction of desired symbol $\mathbf{s}_m(k)$, point $B$ denotes the intersection of $\overrightarrow{AC}$ and the decision boundary, and $\Phi$ denotes half of the green region's angle range, i.e. $\Phi = \frac{\pi}{M}$.
We can observe that when the received signals set in the green region, it can be correctly detected.
Moreover, the distance between point $B$ and point $C$ reflects the received signal's robustness to the disruption of the noise.
Therefore, $\left|\overrightarrow{BC}\right|$ is an measurement to evaluate the SER, i.e. the larger $\left|\overrightarrow{BC}\right|$, the lower SER.
After rotating the diagram clockwise by $\angle{\mathbf{s}_m(k)}$ degrees as in Fig. \ref{fig:CR2}, $\left|\overrightarrow{BC}\right|$ can be easily expressed as
\begin{equation}
\sqrt{\frac{P}{N}}\left[\mathfrak{R}\left\{\mathbf{h}_k^H \bm{\theta}_m e^{-j\angle{\mathbf{s}_m(k)}}\right\}\tan\Phi
-\left|\mathfrak{I}\left\{\mathbf{h}_k^H \bm{\theta}_m e^{-j\angle{\mathbf{s}_m(k)}}\right\}\right|\right].
\end{equation}
Since our goal is to minimize the maximum SER of all users, we can use $\left|\overrightarrow{BC}\right|$ in (4) as the SER metric and write the objective function as
\begin{equation}
\begin{aligned}
&\underset{\bm{\theta}_m}{\max}~~\underset{k}{\min}~~ \left|\overrightarrow{BC}\right| \\
&~~~~~~~~\mathrm{s.t.}~~~\bm{\theta}_m(n) \in \mathcal{F}, \forall n.
\end{aligned}
\end{equation}
For simplicity, we will ignore the $\sqrt{\frac{P}{N}}$ term in the follows, which has no influence on the objective function.
Then, the SER minimization problem can be reformulated as
\begin{subequations}
\label{eq:p1}
\begin{align}
\label{eq:p1 func}
&\underset{\bm{\theta}_m}{\min}~~\underset{k}{\max}~~\left|\mathfrak{I}\left\{\widetilde{r}_{k,m}\right\}\right|-\mathfrak{R}\left\{\widetilde{r}_{k,m}\right\}\tan \Phi \\
&~~~~~~~~\mathrm{s.t.}~~~\widetilde{r}_{k,m} = \mathbf{h}_k^H \bm{\theta}_m e^{-j\angle{\mathbf{s}_m(k)}}, \forall k,\\
\label{eq:p1 theta}
&~~~~~~~~~~~~~~~\bm{\theta}_m(n) \in \mathcal{F}, \forall n,
\end{align}
\end{subequations}
which is a NP-hard problem due to the non-convex constraint (\ref{eq:p1 theta}).
Considering the unaffordable computational complexity of exhaustive search, we will propose an efficient suboptimal solution in the next section.
\section{Symbol-level Precoder Design for Low Resolution Reflecting Elements}
\vspace{0.2 cm}
\label{sec:MISO}
In this section, we attempt to solve (\ref{eq:p1}) with a suboptimal method, which first solves (\ref{eq:p1}) with the low-resolution variables being relaxed into their continuous counterparts, and then finds the nearest discrete phase values to the obtained optimal solutions.

The SER minimization problem with infinite resolution $\bm{\theta}_m$ can be expressed as
\begin{subequations}
\label{eq:p1 continue 7}
\begin{align}
\label{eq:p1 continue 7a}
&\underset{\bm{\theta}_m}{\min}~~\underset{k}{\max}~~\left|\mathfrak{I}\left\{\widetilde{r}_{k,m}\right\}\right|-\mathfrak{R}\left\{\widetilde{r}_{k,m}\right\}\tan \Phi \\
&~~~~~~~~\mathrm{s.t.}~~~\widetilde{r}_{k,m} = \mathbf{h}_k^H \bm{\theta}_m e^{-j\angle{\mathbf{s}_m(k)}}, \forall k,\\
&~~~~~~~~~~~~~~~\left|\bm{\theta}_m(n)\right| = 1, \forall n.
\end{align}
\end{subequations}
In order to facilitate the algorithm development, we convert the complex-valued problem into real-valued problem by denoting
\begin{equation}
\begin{aligned}
&\mathbf{\widetilde{h}}_{k,m}^R \triangleq \mathfrak{R}\left\{\mathbf{h}_k e^{j\angle{\mathbf{s}_m(k)}}\right\}, \;\;\;
\mathbf{\widetilde{h}}_{k,m}^I \triangleq \mathfrak{I}\left\{\mathbf{h}_k e^{j\angle{\mathbf{s}_m(k)}}\right\}, \\
&\bm{\theta}_m^R \triangleq \mathfrak{R}\left\{\bm{\theta}_m\right\}, \;\;\;
\bm{\theta}_m^I \triangleq \mathfrak{I}\left\{\bm{\theta}_m\right\},\;\;\;
\bm{\Theta}_m \triangleq [\bm{\theta}_m^R,\bm{\theta}_m^I]^T.
\end{aligned}
\end{equation}
Then, letting $\alpha = \tan\Phi$, (\ref{eq:p1 continue 7a}) can be expressed as
\begin{equation}
\underset{\bm{\theta}_m}{\min}~~\underset{i}{\max}(g_{2i-1},g_{2i}),
\end{equation}
where
\begin{equation}
\begin{aligned}
&g_{2i-1} \triangleq \left(\mathbf{\widetilde{h}}_{k,m}^I-\mathbf{\widetilde{h}}_{k,m}^R\alpha\right)^T\bm{\theta}_m^R
+\left(\mathbf{\widetilde{h}}_{k,m}^R+\mathbf{\widetilde{h}}_{k,m}^I\alpha\right)^T\bm{\theta}_m^I,\\
&g_{2i} \triangleq -\left(\mathbf{\widetilde{h}}_{k,m}^I+\mathbf{\widetilde{h}}_{k,m}^R\alpha\right)^T\bm{\theta}_m^R
+\left(\mathbf{\widetilde{h}}_{k,m}^I\alpha-\mathbf{\widetilde{h}}_{k,m}^R\right)^T\bm{\theta}_m^I.
\end{aligned}
\end{equation}
Therefore, (\ref{eq:p1 continue 7a}) can be further expressed as
\begin{subequations}
\label{eq:p1 continue}
\begin{align}
\label{eq:p1 continue func}
&\underset{\bm{\Theta}_m}{\min}~~\underset{k}{\max}~~g_i \\
\label{eq:theta mani}
&\mathrm{s.t.}~~~[\bm{\Theta}_m(:,n)]^T\bm{\Theta}_m(:,n)=1,n=1,2,\ldots,N,
\end{align}
\end{subequations}
where $i = 1,2,\ldots,2K$.
Since (\ref{eq:theta mani}) forms a $2N$-dimensional oblique manifold, the optimization problem (\ref{eq:p1 continue}) can be converted into a unconstrained problem in the Riemannain space and further be tackled by some efficient algorithms, e.g. the RCG algorithm \cite{Boumal manopt 14}.
Considering the non-differentiability of the objective function in (\ref{eq:p1 continue func}), the popular smooth log-sum-exp upper bound for the max function is utilized to relax it into
\begin{subequations}
\label{eq:p1 RCG}
\begin{align}
&\underset{\bm{\Theta}_m}{\min}~~\varepsilon\log\left(\sum_{i=1}^{2K}\exp(g_i/\varepsilon)\right) \\
&~\mathrm{s.t.}~~[\bm{\Theta}_m(:,n)]^T\bm{\Theta}_m(:,n)=1,n=1,2,\ldots,N,
\end{align}
\end{subequations}
where $\varepsilon$ is a small positive number.
The optimal continuous phase $\widetilde{\bm{\theta}}_m^*$ can be obtained by the RCG algorithm.
Then, we seek the nearest discrete phase value of $\bm{\theta}_m(n)$ by solving
\begin{subequations}
\label{eq:quantize}
\begin{align}
&\underset{\bm{\theta}_m(n)}{\min}~~~\left|\angle{\widetilde{\bm{\theta}}_m^*(n)}-\angle{\bm{\theta}_m(n)}\right| \\
&~~~\mathrm{s.t.}~~~\bm{\theta}_m(n) \in \mathcal{F}.
\end{align}
\end{subequations}
By (\ref{eq:quantize}), the discrete angle of $\bm{\theta}_m(n)$ can be readily calculated as
\begin{equation}
\label{eq: theta mn}
\angle{\bm{\theta}_m(n)} = \left[\frac{\angle{\widetilde{\bm{\theta}}_m^*(n)}}{\Delta}\right] \times \Delta,
\end{equation}
where $\Delta \triangleq \frac{2\pi}{2^B}$, and $\left[\cdot\right]$ indicates the round operation.

Unfortunately, for the IRS with 1-bit resolution reflecting elements, the quantization error cannot be neglected.
In order to find a better solution, we decompose the precoder $\bm{\theta}_m$ into
\begin{equation}
\bm{\theta}_m = \mathbf{Q}_m \mathbf{v},
\end{equation}
where the auxiliary vector $\mathbf{v} \triangleq [1,-1]^T$ contains all the possible values of 1-bit resolution reflecting elements, $\mathbf{Q}_m \in \left\{0,1\right\}^{N\times2}$ and $\mathbf{Q}_m(n,q)=1$ indicates the $n$-th element in $\bm{\theta}_m$ is $\mathbf{v}(q)$.
Then, the optimization problem (\ref{eq:p1}) can be expressed as
\begin{subequations}
\vspace{-0.2 cm}
\label{eq:p1 1bit}
\begin{align}
&\underset{\mathbf{Q}_m}{\min}~~\underset{k}{\max}~~\left|\mathfrak{I}\left\{\widetilde{r}_{k,m}\right\}\right|-\mathfrak{R}\left\{\widetilde{r}_{k,m}\right\}\tan \Phi \\
&~~~~~~~~\mathrm{s.t.}~~~\widetilde{r}_{k,m} = \mathbf{h}_k^H \mathbf{Q}_m \mathbf{v} e^{-j\angle{\mathbf{s}_m(k)}}, \forall k,\\
&~~~~~~~~~~~~~~\mathbf{Q}_m(n,1) + \mathbf{Q}_m(n,2)= 1, \forall n,\\
&~~~~~~~~~~~~~~\mathbf{Q}_m(n,q) \in \left\{0,1\right\}, \forall n, \forall q.
\end{align}
\end{subequations}
We can observe that (\ref{eq:p1 1bit}) is an MINLP problem, which can be efficiently solved by the popular branch-and-bound algorithm \cite{branch and bound}.
The details of this algorithm are omitted due to the page limitation.
When the optimal $\mathbf{Q}_m^*$ for problem (\ref{eq:p1 1bit}) is found, the optimal precoder $\bm{\theta}_m^*$ can be constructed as
\begin{equation}
\bm{\theta}_m^* = \mathbf{Q}_m^* \mathbf{v}.
\end{equation}
Moreover, we can easily extend this algorithm into the $B$-bit resolution scenario, where the dimension of $\mathbf{Q}_m$ is $N \times 2^B$.
However, considering the computational complexity, the branch-and-bound algorithm is not affordable for high resolution $\bm{\theta}_m$.
\section{Precoder and Combiner Design for Multi-antenna Receivers}
\vspace{0.2 cm}
\label{sec:MIMO}
In the previous section, the symbol-level precoders $\bm{\theta}_m, m = 1,2,\ldots,M^K$, are designed to minimize SER of $K$ single-antenna receivers.
In this section, we consider the multi-antenna receivers, which can properly exploit the combiner to further improve their SER performance.
In an effort to cope with the joint precoder and combiner design problem, we propose an iterative method to decompose and transform it into some tractable sub-problems.

Consider a multiuser multiple-input multiple-output (MU-MIMO) system, where the IRS has $N$ passive reflecting elements to serve $K$ $N_\mathrm{r}$-antenna users.
We denote the combiner at the $k$-th user as $\mathbf{w}_k \in \mathbb{C}^{N_\mathrm{r} \times 1}$ and the channel from IRS to the $k$-th user as $\mathbf{H}_k \in \mathbb{C}^{N \times N_\mathrm{r}}$.
Then, with the desired symbol vector $\mathbf{s}_m$, the received signal at the $k$-th user can be expressed as
\begin{equation}
r_{k,m} = \mathbf{w}_k^H \mathbf{H}_k^H \bm{\theta}_m + n_k.
\end{equation}
We should emphasize that the precoder at IRS is in a symbol-by-symbol fashion, which means that  $\bm{\theta}_m$ varies with the desired symbol vector $\mathbf{s}_m$, but the combiner at the receiver side stays unchanged for all $M^K$ different $\bm{\theta}_m$.
We denote the precoder matrix as $\bm{\Theta} \triangleq \left[\bm{\theta}_1,\bm{\theta}_2,\ldots,\bm{\theta}_{M^K}\right]$,
and the combiner matrix as $\mathbf{W} \triangleq \left[\mathbf{w}_1,\mathbf{w}_2,\ldots,\mathbf{w}_K\right]$.
Similarly, the SER minimization problem can be formulated as
\begin{subequations}
\vspace{-0.1 cm}
\label{eq:p2}
\begin{align}
\label{eq:p2 func}
&\underset{\bm{\Theta},\mathbf{W}}{\min}~~\underset{m}{\max}~~\underset{k}{\max}~~\left|\mathfrak{I}\left\{\overline{r}_{k,m}\right\}\right|-\mathfrak{R}\left\{\overline{r}_{k,m}\right\}\tan \Phi \\
&~~~~~~~~\mathrm{s.t.}~~~\overline{r}_{k,m} = \mathbf{w}_k^H \mathbf{H}_k^H \bm{\theta}_m e^{-j\angle{\mathbf{s}_m(k)}}, \forall k, \forall m,\\
\label{eq:p2 theta}
&~~~~~~~~~~~~~~~\bm{\theta}_m(n) \in \mathcal{F}, \forall n, \forall m, \\
\label{eq:power wk}
&~~~~~~~~~~~~~~\left\|\mathbf{w}_k\right\|^2 \leq 1, \forall k.
\end{align}
\end{subequations}

Since the joint precoder and combiner design in (\ref{eq:p2}) is very difficult to solve, we attempt to obtain $\bm{\Theta}$ and $\mathbf{W}$ in an iterative way by decomposing (\ref{eq:p2}) into the precoder design and the combiner design problems.
With fixed combiner matrix $\mathbf{W}$, (\ref{eq:p2}) is converted into
\begin{subequations}
\label{eq:p2 precoder design}
\begin{align}
&\underset{\bm{\Theta}}{\min}~~\underset{m}{\max}~~\underset{k}{\max}~~\left|\mathfrak{I}\left\{\overline{r}_{k,m}\right\}\right|-\mathfrak{R}\left\{\overline{r}_{k,m}\right\}\tan \Phi \\
&~~~~~~~~\mathrm{s.t.}~~~\overline{r}_{k,m} = \mathbf{w}_k^H \mathbf{H}_k^H \bm{\theta}_m e^{-j\angle{\mathbf{s}_m(k)}}, \forall k, \forall m,\\
&~~~~~~~~~~~~~~~\bm{\theta}_m(n) \in \mathcal{F}, \forall n, \forall m.
\end{align}
\end{subequations}
In (\ref{eq:p2 precoder design}), all $M^K$ precoders $\bm{\theta}_m, \forall m$, are jointly designed, which makes the dimension of variables is relatively high. It is extremely difficult to cope with this problem.
In light of this, we attempt to solve each precoder $\bm{\theta}_m$ by dividing (\ref{eq:p2 precoder design}) into $M^K$ sub-problems.
In the $m$-th sub-problem, $\bm{\theta}_m$ is designed by solving
\begin{subequations}
\vspace{-0.1 cm}
\label{eq:p2 xm}
\begin{align}
&\underset{\bm{\theta}_m}{\min}~~\underset{k}{\max}~~\left|\mathfrak{I}\left\{\overline{r}_{k,m}\right\}\right|-\mathfrak{R}\left\{\overline{r}_{k,m}\right\}\tan \Phi \\
&~~~~~~~~\mathrm{s.t.}~~~\overline{r}_{k,m} = \mathbf{w}_k^H \mathbf{H}_k^H \bm{\theta}_m e^{-j\angle{\mathbf{s}_m(k)}}, \forall k,\\
&~~~~~~~~~~~~~~~\bm{\theta}_m(n) \in \mathcal{F}, \forall n,
\end{align}
\end{subequations}
which has a similar form as (\ref{eq:p1}), and can be solved in the same way as in Sec. \ref{sec:MISO}.

With fixed precoder matrix $\bm{\Theta}$, the combiner design for each receiver is independent.
For the $k$-th receiver, the SER minimization problem can be formulated as
%
\begin{subequations}
\label{eq:p2 wk}
\begin{align}
&\underset{\mathbf{w}_k}{\min}~~\underset{m}{\max}~~\left|\mathfrak{I}\left\{\overline{r}_{k,m}\right\}\right|-\mathfrak{R}\left\{\overline{r}_{k,m}\right\}\tan \Phi \\
&~~~~~~~~\mathrm{s.t.}~~~\overline{r}_{k,m} = \mathbf{w}_k^H \mathbf{H}_k^H \bm{\theta}_m e^{-j\angle{\mathbf{s}_m(k)}}, \forall m,\\
&~~~~~~~~~~~~~~\left\|\mathbf{w}_k\right\|^2 \leq 1,
\end{align}
\end{subequations}
which is a convex problem with $N_\mathrm{r}$ optimized variables, $2 \times M^K$ linear matrix inequality (LMI) constraints and one second-order cone (SOC) constraint.
When using the popular CVX solver \cite{cvx} to solve this problem, the computational complexity under the convergence threshold $\xi$ is $\ln(1/\xi)N_\mathrm{r}\sqrt{2(M^K+1)}(2\times M^K+2N_\mathrm{r}\times M^K+2N_\mathrm{r}^2)$.
The algorithm introduced above is summarized in Algorithm \ref{alg:mimo}, where $N_{\mathrm{max}}$ is the maximum number of iteration, and $\delta_{\mathrm{th}}$ is the threshold to judge convergence.
\begin{algorithm}[!t]
\caption{Precoder and Combiner Design with Multi-antenna Receivers}
\label{alg:mimo}
    \begin{algorithmic}[1]
    \REQUIRE $\mathbf{H}_k$, $\Phi$, $B$, $N_\mathrm{max}$, $\delta_{\mathrm{th}}$.
    \ENSURE $\bm{\Theta}$, $\mathbf{W}$.
        \STATE {Initialize $\bm{\Theta}$, $\mathbf{W}$, $iter=0$, $\delta=\infty$, $f=\infty$.}
        \WHILE {$iter \leq N_{\mathrm{max}}$ and $\delta \geq \delta_{\mathrm{th}}$ }
            \STATE {$f_\mathrm{p} = f$.}
            \FOR {$i=1$ : $M^K$}
                \IF {$B > 1$}
                    \STATE{Obtain $\widetilde{\bm{\theta}}_m^*$ by (\ref{eq:p1 RCG}).}
                    \STATE{Obtain $\bm{\theta}_m(n)$ by (\ref{eq: theta mn}).}
                \ELSE
                    \STATE{Obtain $\bm{\theta}_m(n)$ by (\ref{eq:p1 1bit}).}
                \ENDIF
            \ENDFOR
            \FOR {$k=1$ : $K$}
                \STATE {Obtain $\mathbf{w}_k$ by (\ref{eq:p2 wk}).}
            \ENDFOR
            \STATE {Calculate $f$ = (\ref{eq:p2 func}).}
            \STATE {$\delta = \left|(f-f_\mathrm{p})/f\right|$.}
            \STATE {$iter = iter + 1$.}
        \ENDWHILE
    \end{algorithmic}
\end{algorithm}
\section{Simulation Results}
\vspace{0.2 cm}
In this section, we provide the simulation results to illustrate the effectiveness of our proposed low-resolution symbol-level precoding design algorithms in the MU-MISO and MU-MIMO systems.
We consider a uniform rectangular array (URA) at the IRS, and uncorrelated Rayleigh fading channels between IRS and $K$ receivers.
For simplicity, we assume the channel noise power for all users are the same and $\sigma_k = \sigma = 1, \forall k$.
The SNR is defined as $\frac{P}{\sigma^2}$.
Without loss of generality, the QPSK-modulated signals are assumed in the following simulations.
\begin{figure}[!t]
\centering
  \includegraphics[width=3.5 in]{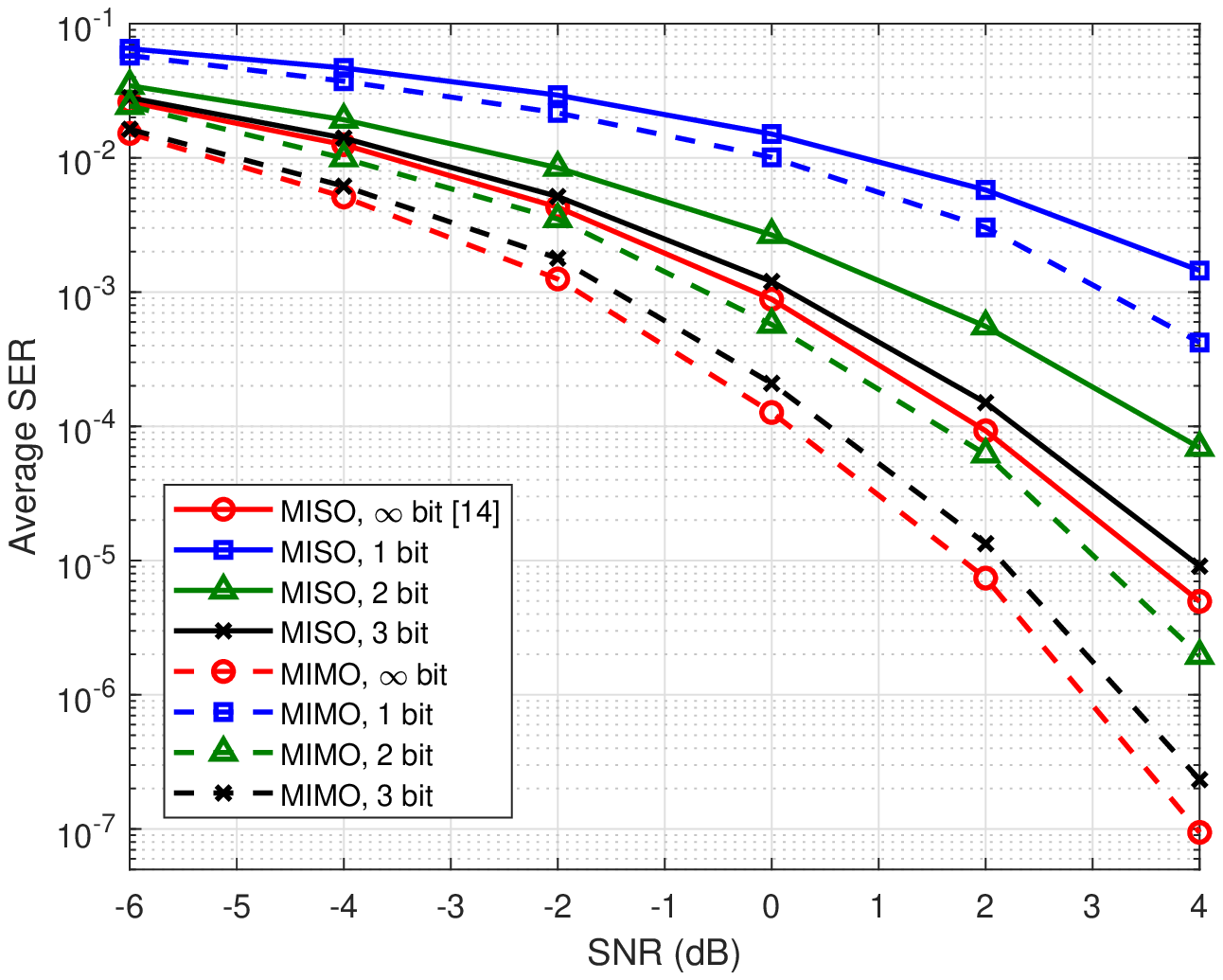}
  \vspace{-0.2 cm}
  \caption{Average SER versus SNR ($K=3$ users, $N=64$ reflecting elements, $N_\mathrm{r}=6$ receive antennas).}
  \label{fig:SER_SNR}
  \vspace{0.5 cm}
  \includegraphics[width=3.5 in]{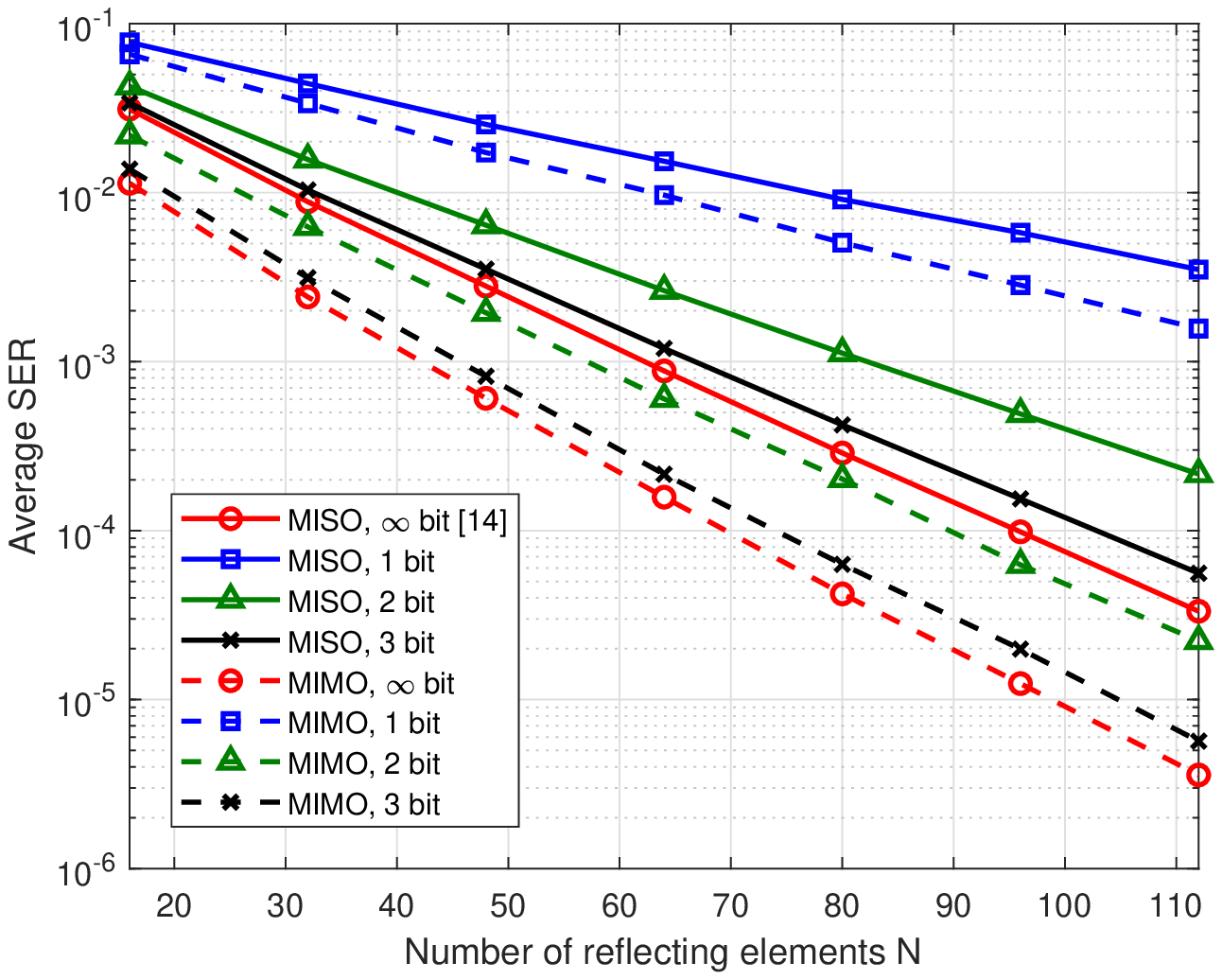}
  \vspace{-0.2 cm}
  \caption{Average SER versus the number of reflecting elements ($K=3$ users, SNR $ = 0$ dB, $N_\mathrm{r}=6$ receive antennas).}
  \label{fig:SER_N} \vspace{-0.3 cm}
\end{figure}
We first present the average SER versus the SNR in Fig. \ref{fig:SER_SNR}, where the IRS equips with $N=64$ reflecting elements to serve $K=3$ users.
For the multi-antenna users, we assume they are all equipped with $N_\mathrm{r}=6$ receive antennas.
The constant envelope precoding design with infinite-resolution IRS in \cite{Liu ISPL 17} is included as a benchmark, where the MU-MISO systems are considered.
In the considered MISO systems (denoted as the solid lines), the infinite-resolution IRS provides the best SER performance because of its flexibility in phase adjustment.
Meanwhile, the 3-bit resolution IRS scheme can achieve almost the same performance as the optimal infinite solution.
However, we observe that when the SNR is high, the 2-bit resolution IRS and 1-bit resolution IRS schemes have 2 dB and 5 dB degradations compared with the optimal solution, respectively.
Therefore, the 3-bit resolution scheme seems a trade-off between the SER performance, energy consumption and the hardware complexity.
We can also arrive at the same conclusion with the multi-antenna receivers.
Moreover, when the users utilize $N_\mathrm{r}$ receive antennas to combine the received signals, the lower SER can be achieved compared with the single-antenna users, which indicates that receive antenna array enables larger processing gain.

In Fig. \ref{fig:SER_N}, we show the average SER versus the number of reflecting elements of IRS when the number of users is $K=3$, the SNR is 0 dB, and the number of receive antennas is $N_\mathrm{r}=6$.
We notice that with the increasing number of reflecting elements, the SER of all schemes decreases, which verifies that the larger IRS can offer the larger beamforming gain.
On the other hand, with larger IRS array, the performance loss due to the low-resolution phases increases as well.
However, the 3-bit resolution IRS scheme can always achieve satisfactory performance with different numbers of reflecting elements.

\begin{figure}[!t]
\centering
  \includegraphics[width=3.6 in]{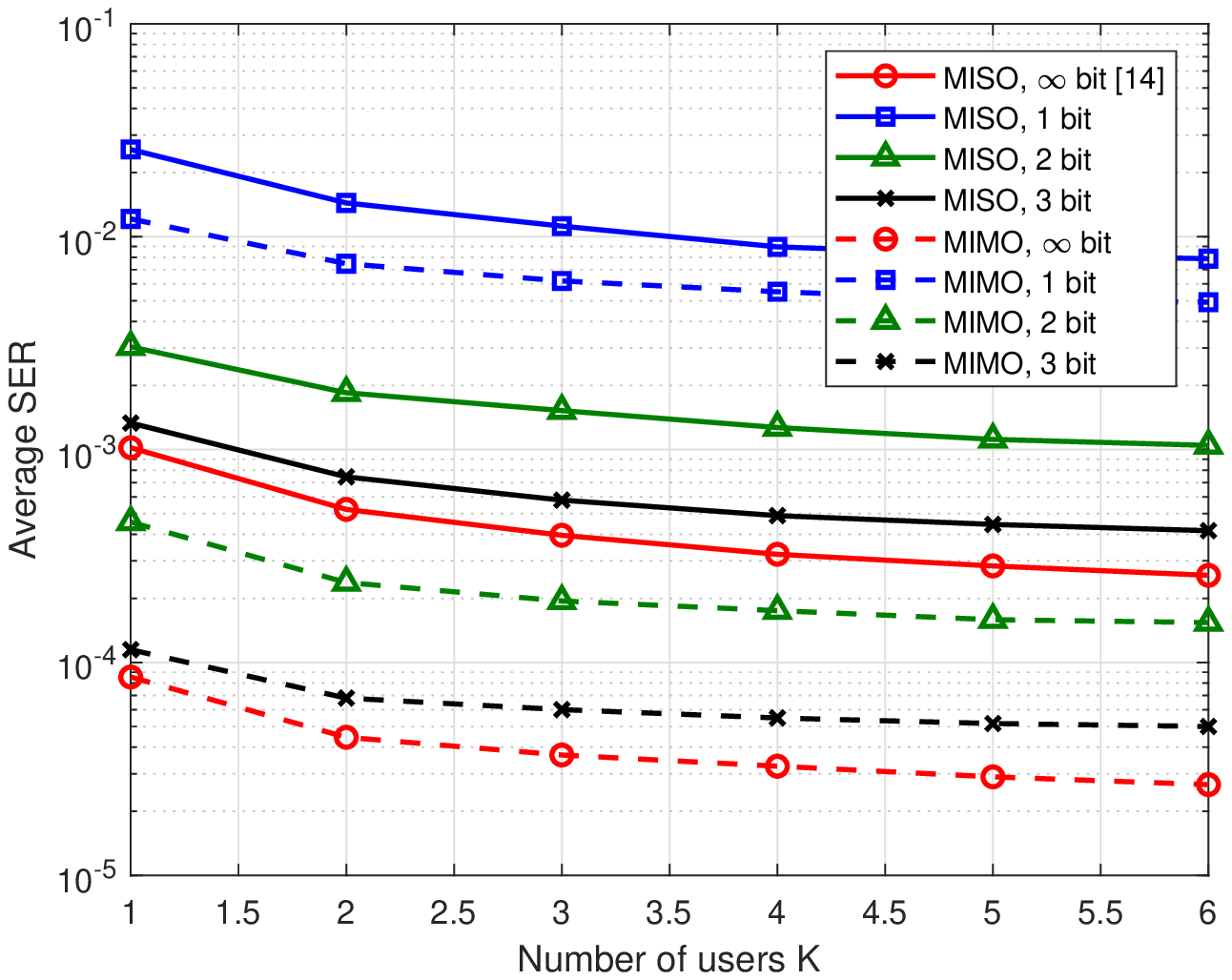}
  \vspace{-0.2 cm}
  \caption{Average SER versus the number of users ($N=64$ reflecting elements, SNR $= -4$ dB, $N_\mathrm{r}=6$ receive antennas).}
  \label{fig:SER_K}
  \vspace{0.5 cm}
  \includegraphics[width=3.6 in]{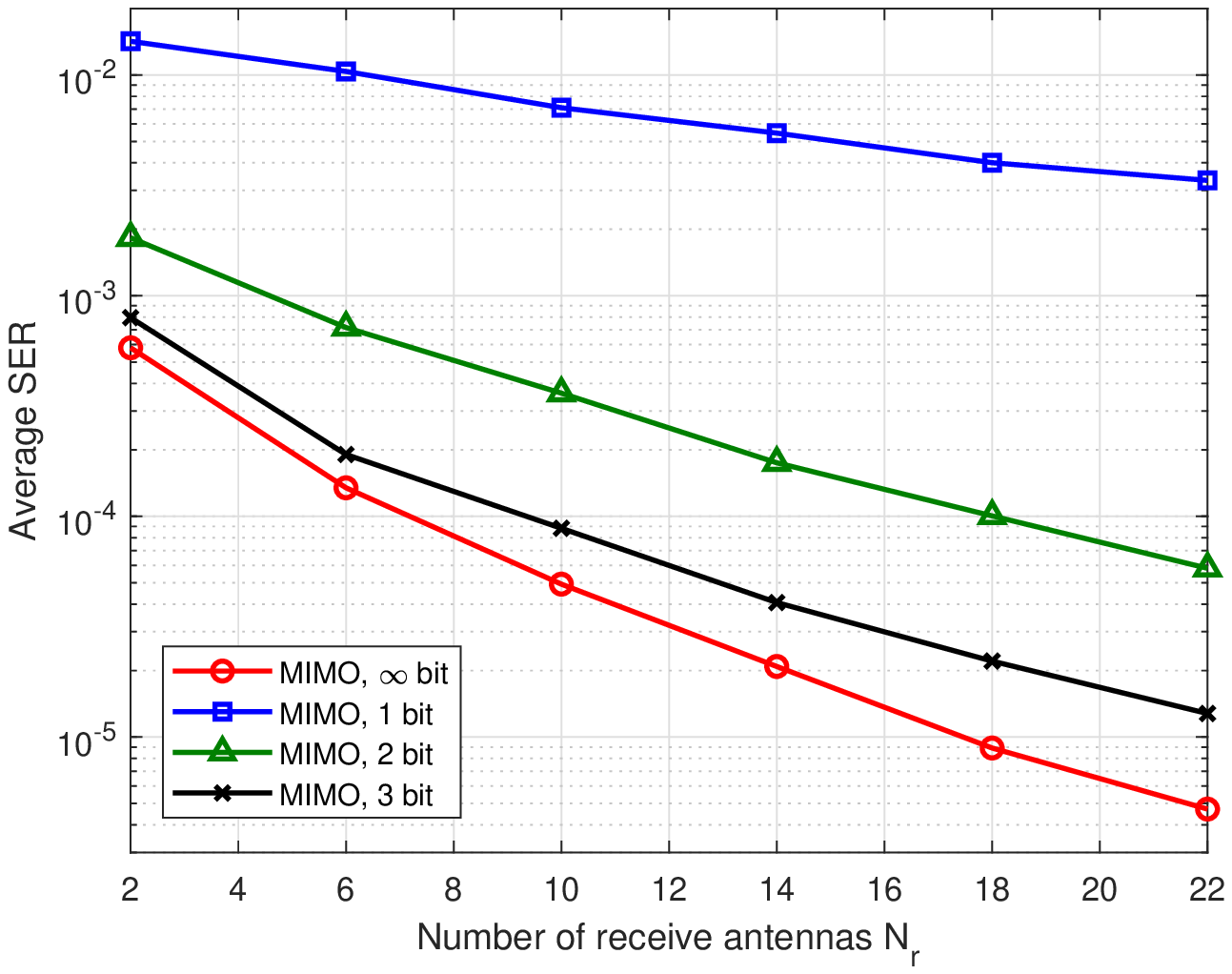}
  \vspace{-0.2 cm}
  \caption{Average SER versus the number of reflecting elements ($K=3$ users, $N=64$ reflecting elements, SNR $= 0$ dB).}
  \label{fig:SER_Nr}\vspace{-0.5 cm}
\end{figure}

Fig. \ref{fig:SER_K} plots the average SER versus the number of users when the number of reflecting elements is $N=64$, the number of receive antennas is $N_\mathrm{r} = 6$, and the transmit power for each user is -4 dB.
The same relationship between different schemes can be observed as in Figs. \ref{fig:SER_SNR} and \ref{fig:SER_N}.
Besides, we can observe that with the increasing number of users, the average SER decreases.
This is because that more users introduce more MUI and the symbol-level precoding scheme exploits MUI to enhance the information transmissions. These results validate the effectiveness of the symbol-level precoding scheme in multi-user systems.

In Fig. \ref{fig:SER_Nr}, we plot the average SER versus the number of receive antennas, where the number of users is $K=3$, the number of reflecting elements is $N=64$, and the SNR is 0 dB.
Since more receive antennas can provide larger processing gain, we see that with the increasing number of receive antennas, the SER of all schemes decreases, which demonstrates the effectiveness of the multiple receive antennas.
\section{Conclusions}
In this paper, we considered the problem that utilizing IRS as a low-complexity, hardware-efficient transmitter in MU-MISO and MU-MIMO systems.
Symbol-level precoding designs for IRS to implement modulation and multiuser transmissions were developed.
The RCG based and branch-and-bound based algorithms were proposed to obtain the low-resolution symbol-level precoder for single-antenna users.
Considering the multi-antenna users, an efficient iterative approach was investigated to design the precoder and combiner iteratively.
Extensive simulation results illustrated the effectiveness of our proposed algorithms.


\begin{thebibliography}{99}
\bibitem{Renzo 19} M. D. Renzo, \textit{et al.}, ``Smart radio environments empowered by AI reconfigurable meta-surfaces: An idea whose time has come,'' \textit{arXiv: 1903.08925}, March 2019.
\bibitem{Basar survey} E. Basar, M. D. Renzo, J. d. Rosny, M. Debbah, M.-S. Alouini, and R. Zhang, ``Wireless communications through reconfigurable intelligent surfaces,'' \textit{IEEE Access}, vol. 7, pp. 116753-116773, Aug. 2019.
\bibitem{Han 08746155} Y. Han, W. Tang, S. Jin, C. Wen, and X. Ma, ``Large intelligent surface-assisted wireless communication exploiting statistical CSI,'' \textit{IEEE Trans. Veh. Technol.}, vol. 68, no. 8, pp. 8238-8242, Aug. 2019. 
\bibitem{Wu 08683145} Q. Wu and R. Zhang, ``Beamforming optimization for intelligent reflecting surface with discrete phase shifts,'' in \textit{Proc. IEEE Int. Conf. Acoust. Speech Signal Process. (ICASSP)}, Brighton, United Kingdom, May 2019, pp. 1-5.
\bibitem{Wu 08647620} Q. Wu and R. Zhang, ``Intelligent reflecting surface enhanced wireless network: Joint active and passive beamforming design,'' in \textit{Proc. IEEE Global Commun. Conf. (GLOBECOM)}, Abu Dhabi, United Arab Emirates, Dec. 2018, pp. 1-5.
\bibitem{Cui IWCT} M. Cui, G. Zhang, and R. Zhang, ``Secure wireless communication via intelligent reflecting surface,'' \textit{IEEE Wireless Commun. Lett.}, to appear. 
\bibitem{Shen ICT} H. Shen, W. Xu, S. Gong, Z. He, and C. Zhao, ``Secrecy rate maximization for intelligent reflecting surface assisted multi-antenna communications,'' \textit{IEEE Commun. Lett.}, vol. 23, no. 9, pp. 1488-1492, Sept. 2019. 
\bibitem{Guo 190507920} H. Guo, Y.-C. Liang, J. Chen, and E. G. Larsson, ``Weighted sum-rate optimization for intelligent reflecting surface enhanced wireless networks,'' \textit{arXiv: 1905.07920}, May 2019.
\bibitem{Huang 180901423} C. Huang, G. C. Alexandropoulos, A. Zappone, M. Debbah, and C. Yuen, ``Energy efficient multi-user MISO communication using low resolution large intelligent surfaces,'' in \textit{Proc. IEEE Globecom Workshops (GC Wkshps)}, Abu Dhabi, United Arab Emirates, Dec. 2018, pp. 1-6.
\bibitem{Chen 08742603} J. Chen, Y.-C. Liang, Y. Pei, and H. Guo, ``Intelligent reflecting surface: A programmable wireless environment for physical layer security,'' \textit{IEEE Access}, vol. 7, pp. 82599-82612, June 2019.
\bibitem{Basar IRS-AP 2019} E. Basar, ``Transmission through large intelligent surfaces: A new frontier in wireless communications,'' in \textit{European Conf. Netw, Commun. (EuCNC 2019)}, Valencia, Spain, June 2019, pp. 1-6. 
\bibitem{Masouros ITWC 09} C. Masouros and E. Alsusa, ``Dynamic linear precoding for the exploitation of known interference in MIMO broadcast systems,'' \textit{IEEE Trans. Wireless Commun.}, vol. 8, no. 3, pp. 1396-1404, March 2009.
\bibitem{Li ITWC 18} A. Li and C. Masouros, ``Interference exploitation precoding made practical: Optimal closed-form solutions for PSK modulations,'' \textit{IEEE Trans. Wireless Commun.}, vol. 17, no. 11, pp. 7661-7676, Nov. 2018.
\bibitem{Liu ISPL 17} F. Liu, C. Masouros, P. V. Amadori, and H. Sun, ``An efficient manifold algorithm for constructive interference based constant envelope precoding,'' \textit{IEEE Signal Process. Lett.}, vol. 24, no. 10, pp. 1542-1546, Sept. 2017.
\bibitem{Boumal manopt 14} N. Boumal, B. Mishra, P.-A. Absil, and R. Sepulchre, ``Manopt, a MATLAB toolbox for optimization on manifolds,'' \textit{The J. Mach. Learn. Res.}, vol. 15, no. 1, pp. 1455-1459, 2014.
\bibitem{branch and bound} H. Li, Q. Liu, Z. Wang, and M. Li, ``Transmit antenna selection and analog beamforming with low-resolution phase shifters in mmWave MISO systems,'' \textit{IEEE Commun. Lett.}, vol. 22, no. 9, pp. 1878-1881, July 2018.
\bibitem{cvx} A. Ben-Tal and A. Nemirovski, \textit{Lectures on Modern Convex Optimization: Analysis, Algorithms, and Engineering Applications.}, Philadelphia, USA: Society for Industrial and Applied Mathematics, 2001.

\end{thebibliography}
\end{document}